\documentclass[journal]{vgtc}                     

\onlineid{0}

\vgtccategory{Research}

\vgtcpapertype{application/design study}

\newcommand{\Lumi}{Lumi}

\title{Rhythms of Recovery: Patient-Centered Virtual Reality Exergame for Physical Rehabilitation in the Intensive Care Unit}

\author{%
  Sangjun Eom, 
  Tianyi Hu,
  Wenyi Xu,
  Liheng Zou,
  Ernesto Escobar,
  Gabriel Streisfeld,
  Anna Mall,
  Bradi Granger, \\ and
  Maria Gorlatova
}

\authorfooter{
    \item
  	Sangjun Eom is with Duke University.
  	E-mail: sangjun.eom@duke.edu
    \item
  	Tianyi Hu is with Duke University.
  	E-mail: tianyi.hu@duke.edu

    \item Wenyi Xu is with Duke University.
  	E-mail: ivory.xu@duke.edu

    \item Liheng Zou is with Duke University.
  	E-mail: liheng.zou@duke.edu

    \item Ernesto Escobar is with Duke University.
  	E-mail: ernesto.escobar@duke.edu 

    \item Gabriel Streisfeld is with Duke University.
  	E-mail: gabriel.streisfeld@duke.edu 

    \item Anna Mall is with Duke Heart Center at Duke Hospital.
  	E-mail: anna.mall@duke.edu 

    \item Bradi Granger is with Duke University.
  	E-mail: bradi.granger@duke.edu 

    \item Maria Gorlatova is with Duke University.
  	E-mail: maria.gorlatova@duke.edu 
    }

\abstract{%
    Early mobilization is a structured protocol designed to facilitate motor recovery in intensive care unit (ICU) patients with ICU-acquired weakness. This process is typically implemented by an interdisciplinary team of nurses, physical therapists, and other healthcare professionals. However, its application is often constrained by the patients’ critical conditions, limited mobility, and the challenges of coordinating care within resource-intensive ICU environments. In this study, we developed a patient-centered virtual reality (VR) exergame through an interdisciplinary design process involving clinicians and therapists, tailored to the constraints of critical care. The exergame incorporates progressive mobility levels that mirror early mobilization practices, and includes an embodied avatar to provide guidance and motivation. Using Meta Quest 3 body tracking, the system captures and visualizes patients’ movements, thereby providing motivational engagement and quantifiable mobility metrics. We evaluated the exergame in two stages: a dual-user study involving healthy participants and healthcare professionals or students (N = 13), and a subsequent study with cardiothoracic ICU patients (N = 18) to assess feasibility, design validity, and clinical acceptance. Across both studies, participants reported high enjoyment and engagement without discomfort or stress. Furthermore, patients demonstrated increases in movement speed, range of motion, and workspace volume of the upper body across game levels. Physiological monitoring further indicated that the exergame elicited exertion without inducing excessive cardiovascular responses. These findings highlight the feasibility of VR exergames as a clinically acceptable and engaging adjunct to early mobilization in critical care, offering a novel pathway to improve rehabilitation outcomes for ICU patients. 
}


\teaser{
  \centering
  \includegraphics[width=0.92\linewidth]{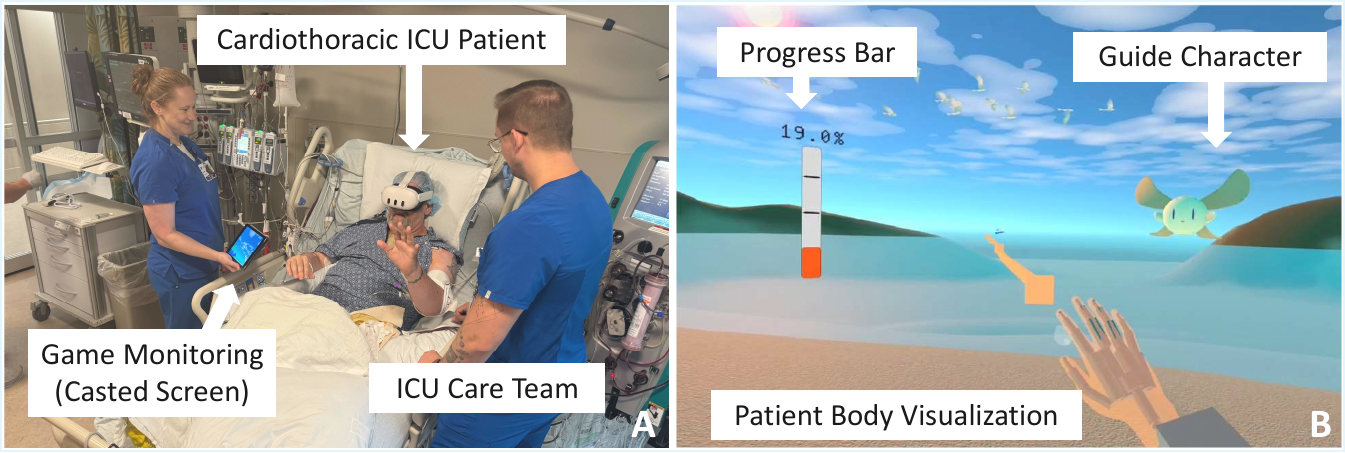}
  \vspace{-0.3cm}
  \caption{A user study setup in the cardiothoracic ICU with a patient recovering from a heart transplant, playing the VR exergame while the ICU care team carefully monitors patients' vital signs and game progress on a casted screen (a). A gameplay scene from one of the exergame levels, featuring a visualization of the patient's upper body, a guiding character, and a progress bar for the patient's progress (b).}
  \label{fig:teaser}
}

\graphicspath{{figs/}{figures/}{pictures/}{images/}{./}} 

\usepackage{tabu}                      
\usepackage{booktabs}                  
\usepackage{lipsum}                    
\usepackage{mwe}                       
\usepackage{amsmath}
\usepackage{mathptmx}                  

\begin{document}


\firstsection{Introduction}

\maketitle

In recent years, virtual reality (VR) has emerged as a promising tool for rehabilitation, offering immersive and interactive environments that can transform conventional therapies into engaging, patient-centered experiences \cite{howard2017meta, fu2022systematic}. In particular, VR exergames, gamified virtual environments designed to promote physical activity, have demonstrated effectiveness in motivating patients to exercise while improving both physical and cognitive functions for therapeutic outcomes \cite{liao2025focus}. Prior research has explored the application of VR exergames in populations such as patients with Parkinson’s disease \cite{marotta2022integrating, chen2020immersive} and elderly individuals with motor dysfunction \cite{palaniappan2018developing, rojo2023pedaleovr}. Despite this growing interest, the use of VR exergames in critical care environments, such as intensive care units (ICUs), remains largely underexplored \cite{mall2024virtual}.

ICU patients frequently experience prolonged immobilization, leading to severe physical deconditioning, neuromuscular impairments, and long-term functional limitations~\cite{soderberg2022fear}. These complications contribute to ICU-acquired weakness, a condition that progressively diminishes both physical and cognitive capacities \cite{vanhorebeek2020icu}. To mitigate these effects, ICU care teams, comprising nurses, physical therapists, and occupational therapists, implement early mobilization protocols, which are tailored to patients’ abilities and clinical stability. While effective in principle, these interventions face considerable barriers, including patients’ dependence on monitoring devices and intravenous lines, restricted mobility due to critical illness, and difficulty in safely moving patients out of bed. Moreover, quantitative assessment of motor progress is rarely feasible in the ICU, with most evaluations relying on subjective or self-reported data \cite{jang2019pulmonary}.


To address these challenges, we present a patient-centered VR exergame designed to support upper-body motor recovery in ICU patients. Developed in interdisciplinary collaboration with clinicians and therapists, the system reflects both therapeutic goals and clinical workflow considerations in the ICU. In the game, patients move their hands in rhythm with music to intercept oncoming visual targets, creating a playful yet structured therapeutic activity. The exergame situates patients in a vibrant, nature-inspired environment to promote positive affect and incorporates a guiding character that embodies a clinician by providing encouragement and fostering a sense of social interaction. Using Meta Quest 3 body tracking, the system captures movements of upper-body joints (shoulder to hand), enabling visualization of motor activity across varying levels of mobility challenges. To ensure patient safety, clinicians can monitor gameplay and movement progress in real time via an external display through Quest 3 screen casting. To our knowledge, this is the first immersive VR system for physical rehabilitation evaluated with ICU patients that integrates embodied interaction, real-time clinician monitoring, and quantitative upper-body movement sensing within routine critical care.

The primary objective of our evaluation was to validate the feasibility, safety, and clinical acceptance of the exergame design under the physical and operational constraints of the ICU. We evaluated the system in two stages: (1) a dual-user study \mbox{(N = 13)} involving healthy participants, and healthcare professionals or students to validate design feasibility and iteratively refine the exergame based on user feedback, and (2) a clinical study (N = 18) with cardiothoracic ICU patients at our university-affiliated hospital to evaluate system feasibility, acceptance, and integration into ICU care workflows. Our research contributions are as follows: 

\begin{itemize}
    \item We developed a patient-centered VR exergame for the ICU co-designed by clinicians, physical therapists, and game designers. The system integrates progressive mobility challenges aligned with early mobilization protocols, embodied guidance via a virtual agent, and clinical monitoring to support safe and engaging upper-body motor recovery.
    \item We conducted a two-stage evaluation: (1) a dual-user study with healthy participants and clinicians (N = 13) to validate feasibility and iteratively refine the design, and (2) a clinical study with cardiothoracic ICU patients (N = 18) and the ICU care team, demonstrating both patient acceptance and clinical integration in a real-world critical care setting. ICU patients reported high enjoyment in physical activity (mean PACES score = 101.9) and positive game experiences, while the ICU care team endorsed the system as easy to use, safe, and effective for bedside use. 
    
    \item We validate the accuracy of Quest 3’s body tracking for upper-body joints and demonstrate methodology for quantifying motor engagement in an ICU context. The pose error of hand and shoulder joints was consistent across all game levels, with errors below 4 cm. We quantified ICU patients’ engagement and observed progressively larger movements, with increases in movement speed (0.06 to 0.18 m/s), range of motion (0.09 to 0.19 m), and workspace volume (0.01 to 0.05 $m^{3}$) across levels. 
\end{itemize}

The remainder of this paper is organized as follows. We first describe related work in Section \ref{sec:relatedwork}, followed by Section \ref{sec:gamedesign} describing the overall game design including the game environment, mobility challenges, and integration of body tracking. We describe the two user study designs in Section \ref{sec:userstudy} and their results in Section \ref{sec:study_results}. We discuss the insights, limitations and future work in Section \ref{sec:discussion}, followed by the conclusion in Section \ref{sec:conclusion}.

\section{Related Work} \label{sec:relatedwork}
\textbf{VR for Physical Rehabilitation.} VR offers unique benefits for physical rehabilitation by creating immersive, interactive environments that enhance patient motivation, engagement, and compliance with therapy protocols \cite{rose2018immersion}. By gamifying repetitive tasks, VR transforms conventional exercises into playful and rewarding activities to motivate users \cite{wang2022supporting, kern2019immersive}, while providing real-time visual and auditory feedback \cite{hamzeheinejad2021impact} and adaptive difficulty levels to match individual abilities \cite{cmentowski2023never}. These features have been shown to promote motor learning and functional recovery across diverse populations, including stroke survivors practicing upper-limb training \cite{afsar2018virtual, duval2022designing}, patients with Parkinson’s disease improving balance and gait \cite{canning2020virtual}, and older adults enhancing coordination and strength 
~\cite{brazil2024effect, shah2023social}. However, while VR rehabilitation has been extensively studied in neurological and aging populations, its application to the highly constrained ICU setting remains limited, largely due to the unique medical and resource constraints of critical care environments~\cite{he2025effects}. 





\textbf{Rehabilitation in Critical Care.}
Early mobilization is a cornerstone of ICU rehabilitation, aiming to counteract the ICU-acquired weakness and mitigate the long-term functional decline through progressive physical activity tailored to the patient's condition and capabilities~\cite{singam2024mobilizing, wang2020early}. Yet, implementation of these protocols is difficult due to patients' critical conditions, dependence on life-support devices, and high risk of overexertion~\cite{soderberg2022fear, linke2020early}. VR has been explored in ICUs primarily for psychological support, such as reducing anxiety and stress during critical care \cite{vlake2022intensive}. Recently, studies have begun to explore VR for physical rehabilitation: \cite{eom2025legato} proposed early-stage ICU-centric game design, and \cite{he2025effects} demonstrated that VR can support early mobilization and improve acceptance of rehabilitation. While this research provided encouraging early evidence of feasibility, it does not provide detailed analyses of patient movement or safety under exertion~\cite{li2024role}. Our work extends this line of evidence by embedding progressive mobility challenges and quantifying upper-body motor engagement with Quest 3 body tracking through evaluation with cardiothoracic ICU patients with post-surgical constraints such as pacemakers, chest tubes, and dialysis lines \cite{jacob2021multidisciplinary, ziyaeifard2018effects}.

\textbf{Quantitative Mobility Assessment}
Quantitative mobility assessment is essential for monitoring rehabilitation progress, with metrics such as range of motion and inter-joint coordination widely recognized as indicators of motor recovery~\cite{zhang2015objective, wang2024quantitative}. Yet, in clinical practice, assessment often relies on subjective tools such as Borg’s Rating of Perceived Exertion (RPE)~\cite{williams2017borg}, the Movement Ability Measure~\cite{allen2007validity}, and the International Physical Activity Questionnaire~\cite{craig2003international}. VR rehabilitation systems can provide more objective alternatives, using handheld controllers~\cite{laver2018virtual, mc2023patient, miller2022temporal} or computer-vision-based tracking~\cite{caserman2019real, kaminska2022usability}. However, controller-based input is less suitable for ICU patients with limited grip strength, motivating our use of the Quest 3’s body tracking to capture upper-body joint motion. Building on prior work with kinematic metrics such as movement entropy~\cite{reinhardt2019entropy} and joint trajectories~\cite{capecci2019kimore}, we focus on three clinically interpretable measures: movement speed, range of motion, and workspace volume to capture mobility progress across exergame levels. While recent work has begun to evaluate the accuracy of body tracking in controlled settings~\cite{casile2023quantitative}, its suitability for clinical rehabilitation, particularly in ICU environments, remains underexplored. Our study addresses this gap by characterizing Quest 3 tracking against ground-truth motion capture and demonstrating its potential for ICU-based mobility assessment. 

\section{Exergame Design Overview} \label{sec:gamedesign}

The main objective of our exergame for the patients is to catch the music lines flying towards them by using their hands (Fig.~\ref{fig:teaser}b). The trajectories and speeds of music lines vary based on the game level, with higher levels requiring broader and faster motions. We developed our VR exergame on Unity 2022.3.46f1 for Quest 3. The demo video of our VR exergame can be accessed at this link\footnote{Link to a demo video: \url{https://tinyurl.com/yn76b4rb}}.

\begin{figure*}[t]
  \centering
  \includegraphics[width=0.92\textwidth]{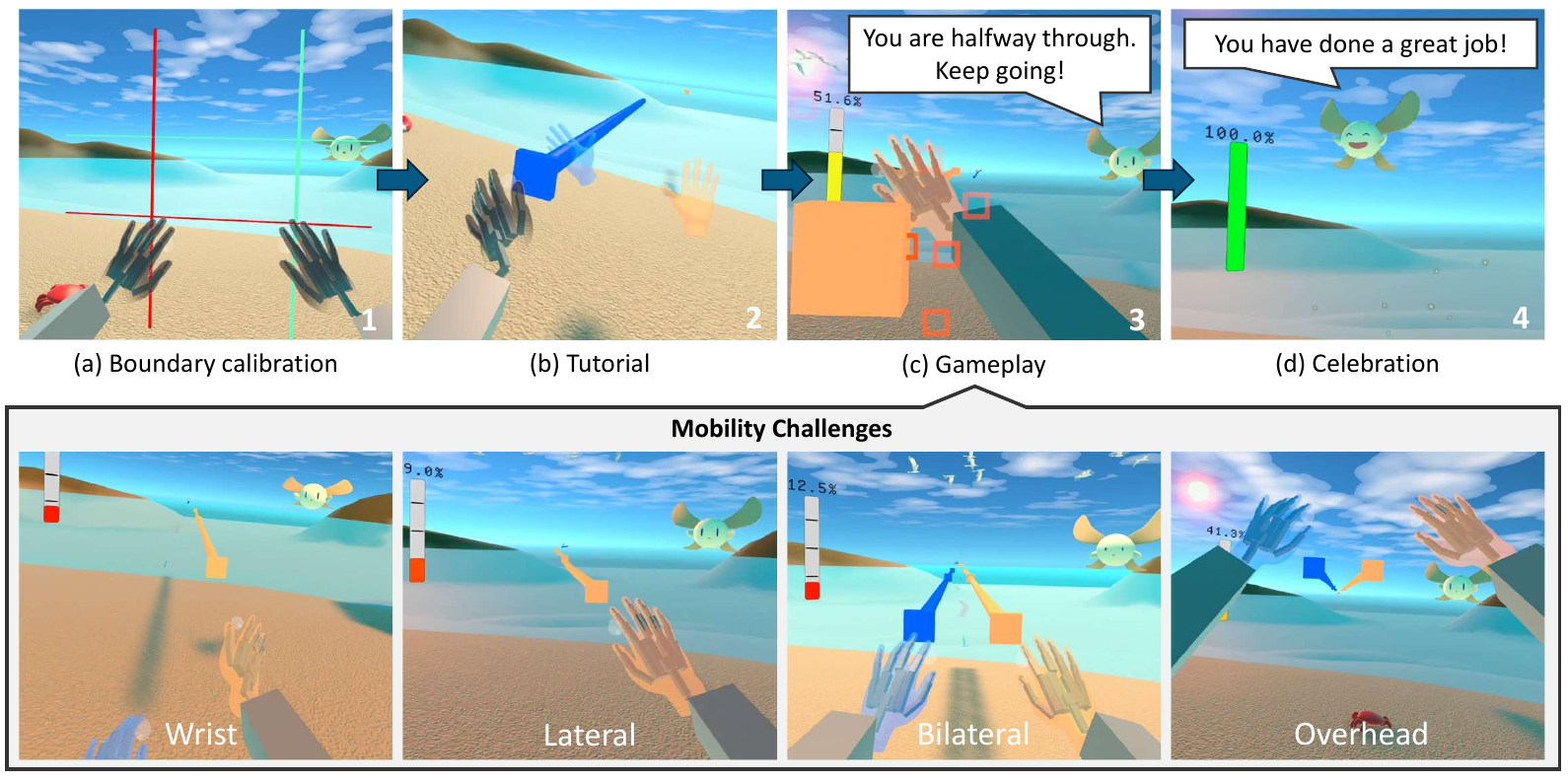}
  \vspace{-0.2cm}
  \caption{The overall flow of the gameplay during our VR exergame: patients setting the maximum range of motion (a), a brief tutorial shown to the patients on how to catch the lines (b), patients playing four game levels consisting of various types of mobility challenges (wrist, lateral, bilateral, and overhead movements) (c), and \Lumi\ celebrating the achievements at the end of each game level (d).}
   \label{fig:gamedesign}
  \vspace{-0.2cm}
\end{figure*}

\subsection{Game Environment and Character}

\label{sec:gameenvironment_character}
ICU environments are often noisy, device-heavy, and visually monotonous, which can contribute to anxiety and depression during prolonged bed rest \cite{saha2022mapping}. 
We selected a calming beach environment for the exergame (Fig.~\ref{fig:teaser}b) that 
incorporates dynamic environmental elements, including a moving crab on the sand and seagulls flying overhead, to enhance immersion and engagement. For User Study I, we maintained consistent lighting and cloud movement speeds across all game levels. However, based on participant feedback, 
we refined the lighting design for User Study II by adjusting the position and trajectory of the virtual sun. Specifically, the light source now starts lower on the horizon and gradually ascends to simulate sunrise, then transitions toward sunset as game levels progress. This enhancement was designed to 
make the environment feel more dynamic and realistic.

The presence and active guidance of clinicians, including monitoring, instruction, and emotional encouragement, are critical to supporting patient engagement and successful rehabilitation outcomes in the ICU setting \cite{tironi2019empathic}. Because VR headsets reduce awareness of the real-world care team, we designed a virtual guide character, \textit{\Lumi} (Fig.~\ref{fig:teaser}b), a fairy-like character who provides instructions, encouragement, and feedback through animated facial and body expressions to embody the clinician’s role. 


Prior research demonstrates that virtual avatars can motivate patients, foster social engagement, and enhance both user experience \cite{graf2023emotional} and task performance \cite{bartl2022effects}. Inspired by these findings, \Lumi\ was designed to be friendly, cheerful, and optimistic, incorporating expressive facial animations, natural gestures, and calm vocal tones to foster a sense of comfort and companionship. During gameplay, \Lumi\ provides step-by-step instructions for each rehabilitation task, offers encouraging feedback when patients reach performance milestones (25\%, 50\%, and 75\% completion), and celebrates their success at the end of each level through distinct animations and positive reinforcement. 

Based on feedback, \Lumi’s design evolved between studies. In User Study I, the character moved dynamically within the scene using Quest’s Passthrough Camera API~\cite{metapassthrough} and a Quest-optimized YOLOv9 model~\cite{metaunitysentis} to align with the clinician's position. This was later simplified to a fixed standing location in \mbox{User Study II} to reduce distraction. Similarly, \Lumi’s initial voice, generated using the Meta Voice SDK, was replaced with more natural text-to-speech audio, generated from Minimax Audio Generator~\cite{minimax}, in User Study II to improve warmth and empathy (see Section~\ref{sec:study1_gameexperience}).

\subsection{Game Progression}

Once the application is initialized, \Lumi\ welcomes the patient and introduces itself. To help the patient familiarize themselves with the tracked hand and body visuals captured by the Quest 3, \Lumi\ prompts the patient for a high five. Following this introduction, \Lumi\ guides the patient through a boundary calibration process (Fig.~\ref{fig:gamedesign}a), instructing the patient to first rest their hands on their laps, then raise their arms as high as possible, and finally extend their arms laterally. These motions establish the patient-specific movement boundaries for the exergame. This calibration step is essential in the ICU setting, as patients exhibit varying degrees of mobility depending on their surgery, recovery stage, and overall condition. After calibration, a brief tutorial is presented, demonstrating how to use hand movements to interact with visual targets, such as touching the virtual lines (Fig.~\ref{fig:gamedesign}b). After the tutorial is demonstrated, the patient plays the exergame from level 1 to level 4 in order (Fig.~\ref{fig:gamedesign}c). Each level lasts two minutes, accompanied by a distinct background music track, and challenging mobility tasks designed to gradually enhance upper-body motor engagement. This duration was set to match the limited endurance of ICU patients, but can be adjusted in future implementations. 

\begin{table}[t]
  \caption{Descriptions of mobility challenges in each game level.}
  \vspace{-0.1cm}
  \label{tab:gamelevel}
  \scriptsize%
  \centering%
  \begin{tabular}{>{\centering\arraybackslash}m{1.0cm} |
                  >{\centering\arraybackslash}m{1.8cm}  
                  >{\centering\arraybackslash}m{1.5cm} 
                  >{\centering\arraybackslash}m{1.5cm}}
  \toprule
   \textbf{Levels} & \textbf{Music Rhythm (BPM)} & \textbf{Movement Types} & \textbf{Duration of Hold (s)} \\
  \midrule
    L1 & 77 & Wrist & 4-6 \\
    L2 & 105 & Lateral & 6-8 \\
    L3 & 112 & Bilateral & 8-10 \\
    L4 & 140 & Overhead & 10-12 \\
  \bottomrule
  \end{tabular}%
  \vspace{-0.6cm}
\end{table}

\subsection{Levels of Mobility Challenges}

Due to the varying degrees of mobility limitations among ICU patients, physical therapists typically provide individualized rehabilitation practices tailored to each patient’s condition. These practices usually begin with simple motor exercises, such as gentle hand and wrist rotations, and gradually progress to movements involving larger joints. As mobility improves, exercises incorporate elbow and shoulder motions to support lateral, overhead, and bilateral arm activities. To ensure that our VR exergame closely mirrored clinical practice, we designed and implemented four progressive levels of mobility challenges in collaboration with an expert physical therapist with over 10 years of clinical experience in the ICU.  

Each game level was developed to gradually increase both range of motion and exertional demand (Tab.~\ref{tab:gamelevel}). Level 1 (L1), set to music at 77 BPM, primarily targets wrist mobility with limited range and short hold durations. Movements included alternating wrist rotations and six-second holds on the left and right sides. Level 2 (L2), at 105 BPM, introduced greater emphasis on unilateral arm movements across larger horizontal and vertical ranges, with hold durations extended to eight seconds. Level 3 (L3), paced at \mbox{112 BPM}, expanded to bilateral arm movements, requiring participants to hold both arms overhead for 10 seconds, alternating between lateral and bilateral motions. Finally, Level 4 (L4), synchronized to a fast rhythm of 140 BPM, incorporated wide diagonal movements that required each arm to cross the body midline, with hold durations increased to 12 seconds. The progression of beats per minute (BPM) across levels was informed by prior research on music-supported exercise, which shows that faster tempos are associated with higher exertional intensity, greater movement synchronization, and increased enjoyment of physical activity \cite{karageorghis2011revisiting, karageorghis2012music}.  


\subsection{Body Tracking and Movement Analysis}

Accurate movement representation of patients' bodies is critical for enabling them to perceive their actions and monitor their progress toward rehabilitation goals in a fully virtual environment \cite{wang2022survey}. While hand-held controllers (e.g., Quest \cite{wang2022supporting} or VIVE \cite{elor2018project, kern2019immersive}) are commonly used to visualize users’ hand movements by tracking the controllers’ motion, this approach is less suitable for ICU patients due to reduced hand strength and limited ability to grasp or hold controllers. To address this limitation, we utilized the built-in upper-body tracking functionality of the Quest 3 \cite{metabodytracking}, allowing real-time visualization of patients’ upper-body movements without requiring external tracking devices (Fig. \ref{fig:teaser}b). Our implementation focused exclusively on upper-body tracking, capturing real-time pose data for the hands (palm), elbows, and shoulders to enable patients to perceive their movements naturally. Joint data were sampled at 50 FPS and stored locally on the Quest 3. These datasets were retrieved for post-experiment analysis to evaluate patients’ movement progress using multiple quantitative metrics.

\begin{figure}[t]
  \centering
  \includegraphics[width=0.89\linewidth]{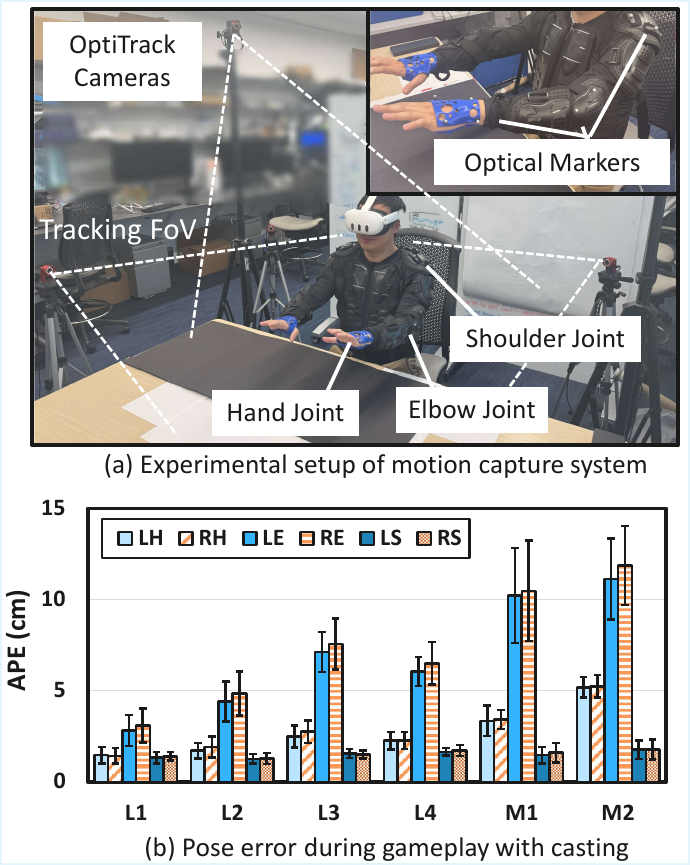}
  \vspace{-0.2cm}
  \caption{Experimental setup of OptiTrack motion capture system for tracking hand, elbow, and shoulder joints (a). APE for each joint (LH/RH: left/right hands, LE/RE: left/right elbows, LS/RS: left/right shoulders) for each game level (L1-L4) and additional motion tasks (M1, M2) with casting (b).}
   \label{fig:tracking_analysis}
  \vspace{-0.3cm}
\end{figure}

\textbf{Evaluation of Tracking Accuracy.}  To ensure the robustness and reliability of the Quest 3’s body tracking during gameplay, 
we assessed the accuracy of upper-body tracking by computing the Absolute Pose Error (APE)~\cite{grupp2017evo} relative to ground-truth motion capture data using eight OptiTrack Flex 3 cameras (57.5 degrees of field-of-view (FoV) lenses), following established methodologies in the motion-tracking literature~\cite{godden2025robotic, hu2024apple}. 
For reference measurements, rigid marker pads containing four infrared markers each were attached to the hands, elbows, and shoulders, with centroids calibrated to anatomical joint positions, across four gameplay levels (L1-L4) and two additional motion tasks involving circular hand movements at slow (M1) and fast (M2) speeds. Motion data were collected from five healthy participants seated upright in a hospital bed with back support, simulating the posture of ICU patients during rehabilitation. To examine the effect of casting, each participant completed the full set of tasks twice: once with screen casting enabled and once without.

As shown in Fig.~\ref{fig:tracking_analysis}b, APEs were lowest at the shoulders (1.25-1.76~cm), moderate at the hands (1.46-5.24 cm), and highest at the elbows (2.82-11.9 cm) during gameplay with casting. Elbow errors increased with both movement distance and speed, reaching their maximum at M2 (left elbow: M = 11.14 cm, SD = 2.22; right elbow: M = 11.87 cm, SD = 2.17). In contrast, mean hand errors remained substantially lower for left hand (M = 5.18 cm, SD = 0.57) and right hand joints (M = 5.24 cm, SD = 0.61). These patterns reflect the fact that Quest~3 directly tracks the hands with onboard cameras, while elbow and shoulder positions are inferred. 

A two-way repeated-measures ANOVA revealed a strong effect of task level on APE, $F(5,20) = 48.92$, $p < .001$, $\eta^2_p = .92$, consistent with prior work showing that tracking error increases with movement speed~\cite{abdlkarim2024methodological}. Despite the highest errors occurring in the additional circular tasks (M1 and M2), APEs within the four exergame levels (L1–L4) remained below 8 cm for elbows and below 4 cm for hands and shoulders, indicating sufficient accuracy for visualizing patient motion. By contrast, the ANOVA showed no significant effect of casting condition, $F(1,4) = 2.80$, $p = .170$, despite a moderate effect size ($\eta^2_p = .41$), demonstrating that enabling screen casting did not degrade Quest 3 tracking accuracy.

\section{User Study Design} \label{sec:userstudy}

To evaluate the effectiveness of our VR exergame in supporting upper-body motor recovery, we conducted two user studies. The first study was implemented in a nursing simulation laboratory with healthy volunteers and recruited healthcare professionals or students. The second study was carried out in a hospital setting with cardiothoracic ICU patients, supported by a clinical team of seven nurses trained in VR applications. Both studies received approval from the Institutional Review Board of Duke University Health System (Protocol: Pro00118509).

\begin{figure}[t]
  \centering
  \includegraphics[width=\linewidth]{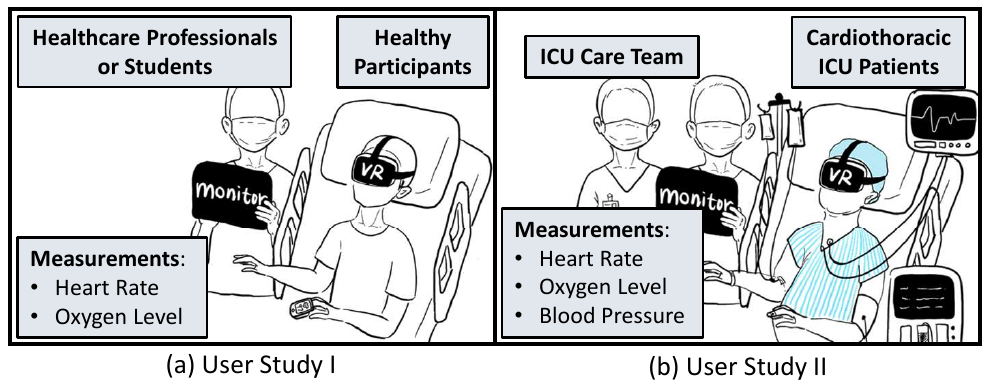}
  \vspace{-0.2cm}
  \caption{User study designs; User Study I, a dual user study consisting of 13 healthcare professionals or students and 13 healthy participants and (a), and User Study II, consisting of the ICU Care team and 18 cardiothoracic ICU patients (b).}
   \label{fig:user_studies}
  \vspace{-0.2cm}
\end{figure}

\subsection{User Study I: Validation of Exergame Design and Feasibility} 

\label{sec:study1}
The primary objective of the first study was to validate the exergame design and assess its feasibility in a simulated hospital environment. The study involved two participant roles: (1) a \emph{patient-role participant}, who engaged in all levels of the mobility challenges provided by the VR exergame, and (2) a \emph{clinician-role participant}, who instructed the patient, supervised the gameplay, and monitored progress via a mirrored external display (Fig.~\ref{fig:user_studies}a).

\subsubsection{Participants}

We recruited 13 healthy participants without upper-body mobility impairments for the patient role (mean age: 25.2 years; range: 19–30 years; 61.5\% male, 38.5\% female) and 13 healthcare professionals for the clinician role (mean age: 28.8 years; range: 23–42 years; 15.4\% male, 84.6\% female). The healthcare professionals were either currently practicing clinicians or students in nursing or physical therapy. In terms of rehabilitation experience, 23.1\% reported no prior experience, 23.1\% reported 0–1 years, 46.2\% reported 2–5 years, and 7.7\% reported over 11 years of experience. Regarding familiarity with VR, 15.4\% of clinicians reported no prior experience, 46.2\% had used VR once or twice, and 38.5\% reported infrequent use (less than once per week). In contrast, among patient-role participants, 7.7\% reported no prior VR experience, 38.5\% had used VR once or twice, 15.4\% used it infrequently, and 38.5\% reported frequent use (more than once per week).


\subsubsection{Procedure}

The experiment was conducted in the nursing simulation laboratory that serves as an accredited training and testing facility for healthcare education and technology evaluation. For our experiment, a hospital bed with adjustable height was prepared in a sitting position, with back support, to accommodate the patient-role participants (Fig.~\ref{fig:user_studies}a). An iPad was used to mirror the VR screen in real time, enabling clinician-role participants to observe gameplay and track patient progress.

Before the experiment, all participants received a briefing and signed informed consent forms, followed by the completion of a pre-experiment questionnaire. Clinician-role participants were then trained on the setup procedure, including launching the VR application on a Quest 3 headset, mirroring the display to an external monitor, and guiding patients through the exergame. To familiarize themselves with the system, clinician-role participants spent approximately 5 minutes wearing the headset and completing the first game level. After this training, the patient-role participants were invited to the simulation bed. The clinician-role participants initiated the VR session, mirrored the display, and instructed the patient-role participants through the gameplay. Patient-role participants completed all four levels of the exergame under clinician supervision.

\subsubsection{Measurements}

\textbf{Patient-Role Participants.} We evaluated the quality of the overall exergame experience using the core module of the Game Experience Questionnaire (GEQ) \cite{ijsselsteijn2013game}. Engagement in mobility challenges was assessed using the Physical Activity Enjoyment Scale (PACES) \cite{kendzierski1991physical}, and the Rate of Perceived Exertion (RPE) Modified Borg Scale \cite{williams2017borg} on a 7-point scale. In addition, participants responded to four items on a 5-point Likert scale assessing the perceived presence and embodiment of the virtual avatar, \emph{\Lumi} (Fig.~\ref{fig:study1_geq}c). At the end of the survey, participants answered open-ended questions regarding aspects they liked and disliked about the exergame. During gameplay, physiological responses were recorded using a pulse oximeter attached to the index finger, measuring heart rate (HR; beats per minute, BPM) and oxygen saturation (SpO\textsubscript{2}; \%) after the completion of each game level.

\textbf{Movement Metrics.} 
To analyze upper-body mobility, we collected pose data, $p$ of 70 upper-body joints, including the hands, tracked by the Quest 3 headset \cite{metabodytracking}, sampled at 50 Hz. From these data, we computed three metrics: (1) speed, (2) range of motion (ROM), and (3) workspace volume.  

The instantaneous speed, $v_i$, of a joint was defined as the Euclidean distance between consecutive positions, $p_i$ and $p_{i-1}$, divided by the elapsed time (0.02s at 50 Hz). The mean speed, $\bar{v}$ over a game level was calculated as,   

\begin{equation}
\setlength{\abovedisplayskip}{-10pt}
\setlength{\abovedisplayshortskip}{-10pt}
\setlength{\belowdisplayskip}{0pt}
\setlength{\belowdisplayshortskip}{0pt}
 \bar{v} = \frac{1}{N-1}\sum_{i=2}^{N} v_i
 \label{eq:speed}
\end{equation}

where $N$ denotes the total number of samples. Speed is reported in meters per second (m/s).  

ROM was quantified as the root-mean-square deviation of joint positions from their centroid, $c$, representing the spatial dispersion of a joint’s trajectory,

\begin{equation}
\setlength{\abovedisplayskip}{-10pt}
\setlength{\abovedisplayshortskip}{-10pt}
\setlength{\belowdisplayskip}{0pt}
\setlength{\belowdisplayshortskip}{0pt}
 ROM = \sqrt{\frac{1}{N} \left\lVert \sum p_i - c \right\rVert ^2}
 \label{eq:rangeofmotion}
\end{equation}

with results reported in meters (m). Larger values indicate greater variability in joint positioning.  

The workspace volume, defined as the three-dimensional space encompassed by all joint positions, was quantified by computing the convex hull of joint trajectories recorded for each game level. This measure, motivated by prior approaches \cite{lourido2024using, skuric2022approximating}, captures the spatial extent of a patient’s upper-body reach and is commonly used in rehabilitation research as an indicator of functional mobility. Using the standard polyhedral volume formula, 

\begin{equation}
\setlength{\abovedisplayskip}{-10pt}
\setlength{\abovedisplayshortskip}{-10pt}
\setlength{\belowdisplayskip}{0pt}
\setlength{\belowdisplayshortskip}{0pt}
V_{\text{workspace}} = \frac{1}{3}\sum_{k=1}^{K} A_k \, d_k
 \label{eq:workspacevolume}
\end{equation}

where $A_k$  is the area of facet, $k$, $d_k$ is the perpendicular distance from the facet centroid to the reference origin, and $K$ is the number of facets. 
Workspace is reported in cubic meters ($m^3$).



\textbf{Clinician-Role Participants.} We evaluated the usability of our VR exergame for physical rehabilitation in terms of setting up, initiating the app, guiding through each level, and monitoring through a mirrored screen. We selected the System Usability Scale (SUS) questionnaire \cite{brooke1996sus} and measured the completion time of system setup from the start of the experiment to initiating the app on Quest 3. Additionally, we asked open-ended questions about what clinician-role participants liked and did not like about the system and the setup procedure. 


\subsection{User Study II: Evaluation with Cardiothoracic ICU Patients in Critical Care Setting} \label{sec:study2}

The second study was conducted in collaboration with the cardiothoracic ICU care team at our university-affiliated hospital (Fig.~\ref{fig:user_studies}b). Over the course of one month, post-surgical ICU patients who provided informed consent were recruited. Each session was conducted in the patient’s room. To ensure patient safety and maintain procedural consistency, a dedicated care team of seven experienced nurses oversaw all 18 sessions.

\subsubsection{Participants}
A total of 18 cardiothoracic ICU patients were recruited (mean age: 54.9 years; range: 31–73 years; 83.3\% male, 16.7\% female; 33.3\% Black or African American, 66.7\% White). The mean duration of ICU stay at the time of participation was 8.4 days (SD = 14.7), with two patients having been admitted for more than 48 days. Surgical procedures among participants included heart transplantation, coronary artery bypass grafting, aortic surgery, and mitral valve replacement. In terms of prior VR exposure, 61.1\% of patients had never used a VR headset, while 38.9\% had used one once or twice.

The ICU care team comprises seven expert nurses (mean age = 37.5 years; range: 25–61; 42.9\% male, 57.1\% female; three with more than 11 years of clinical experience, two with 6–10 years of clinical experience, and two with 2–5 years of clinical experience). All nurses have worked with more than 100 patients in the ICU setting except for one nurse who had worked with fewer than 10 ICU patients.

\subsubsection{Procedure}

The experimental procedure largely followed that of Study I with adaptations for the ICU context. After obtaining informed consent, the care team assisted patients with donning the Quest 3 headset and initiated the VR exergame. The headset display was mirrored to an external monitor, enabling the care team to track patient progress throughout the session. For all patients, the care team read each survey question aloud and either allowed patients to record their responses independently or recorded the responses on their behalf when provided verbally.

\textbf{Changes from User Study I.} The exergame version used in Study~II incorporated several design modifications based on feedback from Study I. Specifically, \emph{\Lumi} was repositioned to improve usability: during calibration, tutorials, and celebratory sequences, \Lumi\ remained centered, while during gameplay, \Lumi\ was placed to the right (Fig.~\ref{fig:gamedesign}). \Lumi’s voice was replaced with a more natural-sounding voice. In addition, environmental dynamics were enhanced by varying the lighting direction and cloud speed across levels, producing distinct visual settings aligned with the rhythm of the accompanying music track.

\subsubsection{Measurements}

Measurements for patient participants were consistent with those in Study I, with the addition of blood pressure monitoring. Physiological signals, including systolic blood pressure (SBP), diastolic blood pressure (DBP), heart rate, and oxygen saturation, were recorded from the ICU bedside monitor at baseline and after each game level. In Study I, the SUS questionnaire was completed by clinicians to evaluate system setup and operation. For Study II, SUS was omitted because the same ICU care team facilitated all patient sessions. Instead, each nurse provided qualitative feedback on system usability through survey-based, open-ended questions, focusing on their experience with setup, facilitation, and patient monitoring.

\section{Study Results} \label{sec:study_results}

Participant responses to open-ended questions are reported with identifiers denoting their study role: \textit{H} for healthy participants in the patient role, and \textit{S} for healthcare professionals or students in the clinician role (User Study I). \textit{P} for ICU patients and \textit{C} for the care team (User Study II). For all analyses of physiological changes and movement performance, one-way repeated-measures ANOVA was used to examine within-subject effects across game levels.  

In User Study II, one patient (P14) reported dizziness and nausea during gameplay due to their critical health condition and therefore completed only the first three game levels; as a result, this patient’s data was excluded from the overall analysis. 

\subsection{User Study I}
\subsubsection{System Usability from Clinician Perspective} 

The average SUS score for clinician-role participants was high (M = 79.0, SD = 12.5; excluding one outlier). 
The mean system setup time was 6.3 minutes (M = 377.5s, SD = 113.7s), suggesting that even clinicians with no prior VR experience were able to cast the VR display to a monitor and successfully operate the game for patient-role participants. This finding aligns with open-ended feedback, where participants noted that the system was “easy to navigate” (S3, S4) and that “instructions were clear” (S12, S13).  

Additionally, clinician-role participants expressed strong enthusiasm for the use of VR in critical care, highlighting its potential for implementation in the ICU. They valued its role as a “new or better way to engage with patients” (S1, S2, S8, S9), its “immersive quality” (S5, S12), and its “visualization of patient body during game” (S13). They further endorsed the system as an “excellent addition to physical rehabilitation” (S5), expressed interest in seeing it deployed in hospitals (S9, S12), and emphasized its “huge potential for very ill patients to engage their upper extremity” (S10, S11).

At the same time, clinician-role participants noted a learning curve in adopting VR technology. The most frequently reported challenges included “setting up for casting” (S2, S7, S9, S11, S12), the “limited FoV of the casted screen” (S10), and “clicking a button on the Quest controller” (S8). These comments suggest that successful deployment in ICU settings will require additional training and support to ensure safety and smooth clinical integration.


\begin{figure}[t]
  \centering
  \includegraphics[width=0.95\linewidth]{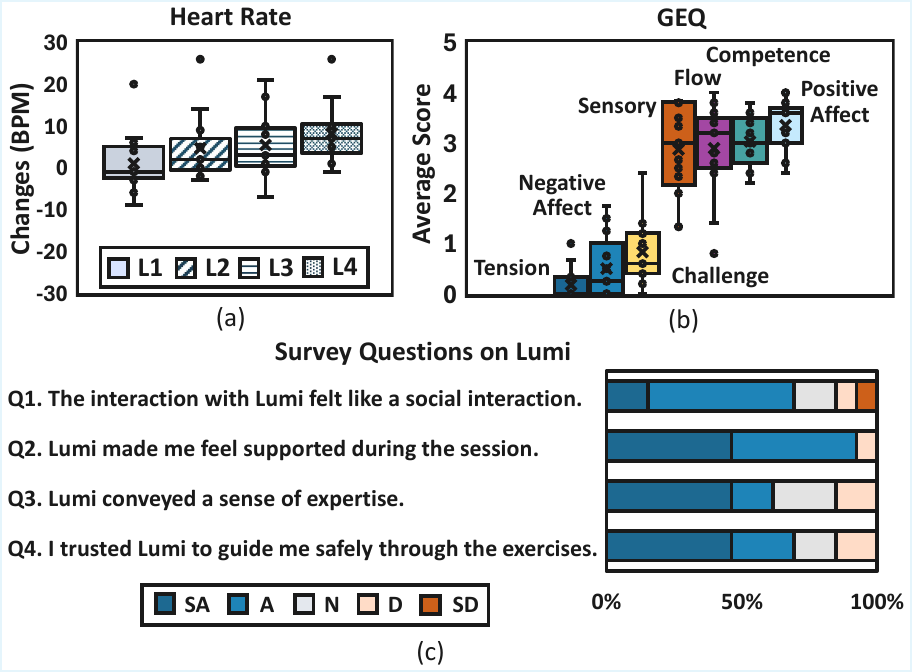}
  \vspace{-0.2cm}
  \caption{User Study I results: percentage changes in healthy participants' heart rates measured in BPM across four game levels, L1-L4 (a), average scores for each category in GEQ (b), and survey questions on \Lumi\ in 5-point Likert scale; SA: strongly agree, A: somewhat agree, N: neither agree nor disagree, D: somewhat disagree, SD: strongly disagree (c).}
   \label{fig:study1_geq}
  \vspace{-0.2cm}
\end{figure}

\begin{figure*}[t]
  \centering
  \includegraphics[width=0.95\textwidth]{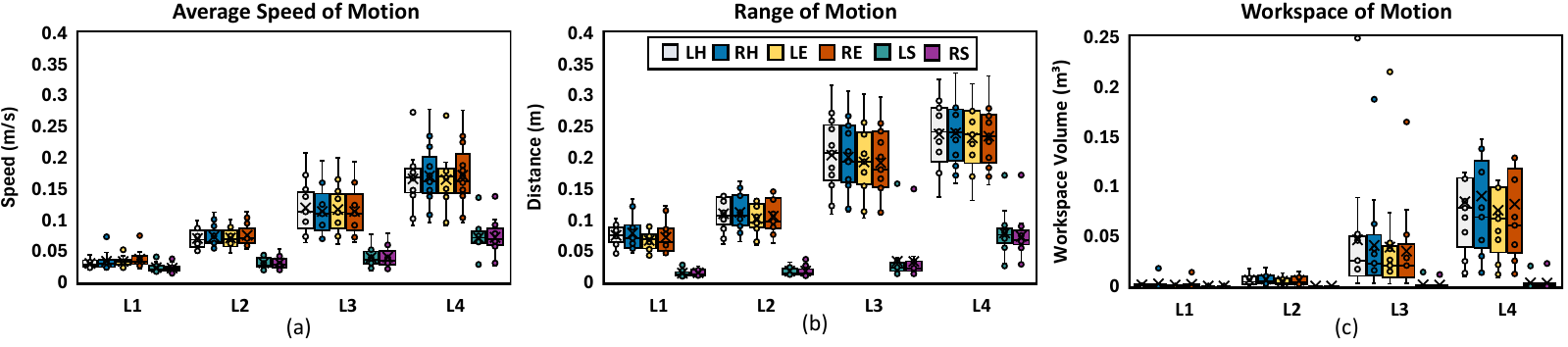}
  \vspace{-0.2cm}
  \caption{User Study I results: healthy participants' motion analysis from Quest 3 body tracking for each body joints (LH/RH: left/right hands, LE/RE: left/right elbows, LS/RS: left/right shoulders) in terms of speed (a), range of motion (b), and workspace volume (c).}
   \label{fig:study1_movement}
  \vspace{-0.4cm}
\end{figure*}

\subsubsection{Game Experience} \label{sec:study1_gameexperience}

The GEQ results indicated high scores for \textit{positive affect} (M = 3.35, SD = 0.48), \textit{competence} (M = 3.03, SD = 0.48), \textit{flow} (M = 2.89, SD = 0.88), and \textit{sensory and imaginative immersion} (M = 2.87, SD = 0.89), on a 0–4 scale, in Fig. \ref{fig:study1_geq}b. These findings suggest that the exergame provided participants with enjoyable, smooth, and immersive experiences while fostering a sense of skillfulness and achievement in completing the mobility challenges. In contrast, low scores were observed for \textit{challenge} (M = 0.83, SD = 0.61), \textit{negative affect} (M = 0.50, SD = 0.60), and \textit{tension} (M = 0.18, SD = 0.31). These results were expected, as the patient-role participants did not have mobility impairments and therefore did not experience the exergame as overly difficult or stressful.  

\textbf{Survey Questions on \Lumi.} We define \textit{positivity rate} as the percentage of responses in the “strongly agree” (SA) and “somewhat agree” (A) categories (Fig. \ref{fig:study1_geq}c). Most participants felt supported by \Lumi\ during the session (positivity rate: 92.3\%). Furthermore, 69.2\% agreed that interaction with \Lumi\ resembled social interaction and that they trusted \Lumi\ to guide them safely through the exercises. A lower proportion (61.5\%) agreed that \Lumi\ conveyed a sense of expertise. These findings align with open-ended feedback, in which participants highlighted \Lumi’s “appropriate instructions” (H1, H7), “calm and clear voice” (H5, H6, H7, H9), and described \Lumi\ as contributing to positive emotions, such as making them feel “happy and relaxed” (H8, H10) or “not alone” (H3).  

Open-ended responses also identified areas for improvement in \Lumi’s design. While participants appreciated \Lumi’s presence for “adding more interactions” (H11), “making the environment more dynamic” (H2), and “reducing feelings of isolation” (H1, H3, H7), some reported that \Lumi’s movement was “obstructing their view” (H1, H2, H11), felt “too close” (H8), or was “distracting” (H11). In response, we modified \Lumi’s behavior so that the character remains static at the side during gameplay and only moves to the center when delivering instructions or celebrating task completion. Participants also noted that \Lumi’s voice sounded “too artificial” (H4, H6) and that its facial expressions lacked dynamism (H3). To address this, we replaced the audio with more natural, human-like recordings and updated facial animations (see Section \ref{sec:gameenvironment_character}).

\subsubsection{Physical Activity}

The mean PACES score (M = 103.1, SD = 9.93) indicated that patient-role participants reported high enjoyment of the physical activity during the exergame. In contrast, the mean RPE score was relatively low (M = 2.62, SD = 1.08 on a 7-point scale), suggesting that participants generally perceived the exertion as light. This outcome was expected given that the patient-role participants did not have mobility or health limitations requiring substantial physical effort. Notably, despite the low perceived exertion, the high PACES scores highlight that participants found the gameplay, catching music lines, and completing diverse upper-body mobility challenges engaging and enjoyable.

\subsubsection{Physiological Changes}

The changes in HR (BPM) showed a gradual increase across game levels with progressively more challenging mobility tasks (Fig. \ref{fig:study1_geq}a). Relative to baseline, HR increased by 1 BPM (SD = 4.74) after L1, 4.69 BPM (SD = 4.67) after L2, 5.46 BPM (SD = 6.33) after L3, and 8.23 BPM (SD = 3.91) after L4. ANOVA revealed a significant effect of game level on HR, $F(4, 48) = 8.32,\ p < .001$, indicating that exertion increased with task difficulty. This result was corroborated by a nonparametric Friedman test, $\chi^2(4) = 28.78,\ p < .001$. 
These results suggest that the exergame successfully imposed scalable exertional demands, even for healthy participants with no mobility limitations, thereby supporting the validity of the designed mobility challenges. In contrast to HR, oxygen saturation remained stable across game levels, with no significant effect observed (ANOVA: $F(4, 48) = 0.85,\ p = .50$; Friedman: $\chi^2(4) = 3.96,\ p = .41$). 


\subsubsection{Movement Analysis}

As game difficulty increased, mean movement speed rose across all joints: hands, forearms, and shoulders (Fig. \ref{fig:study1_movement}a). For the hands and forearms, speed increased consistently from L1 (LH = 0.030 m/s, RH = 0.033 m/s; LE = 0.033 m/s, RE = 0.037 m/s) to L4 (LH = 0.165 m/s, RH = 0.170 m/s; LE = 0.164 m/s, RE = 0.171 m/s). ANOVA revealed a significant effect of game level on speed for all four joints (LH: $F(3,36)=91.31$, RH: $F(3,36)=97.85$, LE: $F(3,36)=94.43$, RE: $F(3,36)=97.75$, all $p<.001$). Post-hoc comparisons with Bonferroni correction confirmed significant differences between every pair of levels ($p<.001$), indicating systematic progression of speed with task difficulty. Although shoulder joints also showed significant increases (LS: 0.023-0.071 m/s; RS: 0.023-0.072 m/s), the magnitude was smaller (LS: $F(3,36)=47.28$, RS: $F(3,36)=43.10$, both $p<.001$), consistent with their shorter displacement. Post-hoc results nevertheless indicated significant differences across all levels ($p<.05$).  

ROM similarly increased with level progression (Fig. \ref{fig:study1_movement}b). Hands and forearms expanded from less than 10 cm at L1 (LH = 0.075 m, RH = 0.078 m; LE = 0.067 m, RE = 0.072 m) to more than 23 cm at L4 (LH = 0.237 m, RH = 0.239 m; LE = 0.231 m, RE = 0.234 m). ANOVA confirmed a strong effect of level (LH: $F(3,36)=125.18$, RH: $F(3,36)=120.39$, LE: $F(3,36)=127.20$, RE: $F(3,36)=122.53$, all $p<.001$), with post-hoc tests showing significant increases across all level pairs ($p<.001$). Shoulders, in contrast, exhibited more modest ROM gains (LS: 0.015-0.079 m; RS: 0.015-0.073 m). While ANOVA indicated significant effects (LS: $F(3,36)=26.94$, RS: $F(3,36)=25.95$, both $p<.001$), post-hoc significance was limited to higher-level contrasts (LS: L1–L2, L1–L4, L2–L4, L3–L4; RS: L1–L4, L2–L4, L3–L4).  

Workspace volume expanded significantly for the hands and elbows but less so for shoulders (Fig. \ref{fig:study1_movement}c). At L1, workspace was below 0.01 $m^3$ for all joints, but increased markedly by L4 (LH = 0.084 $m^3$, RH = 0.090 $m^3$; LE = 0.076 $m^3$, RE = 0.082 $m^3$). ANOVA confirmed significant effects (LH: $F(3,36)=12.86$, RH: $F(3,36)=12.33$, LE: $F(3,36)=12.08$, RE: $F(3,36)=11.65$, all $p<.001$), with post-hoc tests showing significance for most level transitions. Shoulders remained below 0.01 $m^3$ even at L4 (LS = 0.003 $m^3$, RS = 0.003 $m^3$), with weaker effects (LS: $F(3,36)=3.91,\ p=.016$; RS: $F(3,36)=3.19,\ p=.035$). Post-hoc tests indicated significance only for LS (L3–L4), with no significant contrasts for RS.  

Overall, these results indicate that the exergame systematically scaled movement speed, ROM, and workspace, particularly for the hands and forearms (i.e., elbows). The smaller gains observed for the shoulders reflect their inherently shorter displacement paths, but significant late-level differences suggest that even proximal joints benefited from progressive mobility challenges.  

\begin{figure}[t]
  \centering
  \includegraphics[width=0.94\linewidth]{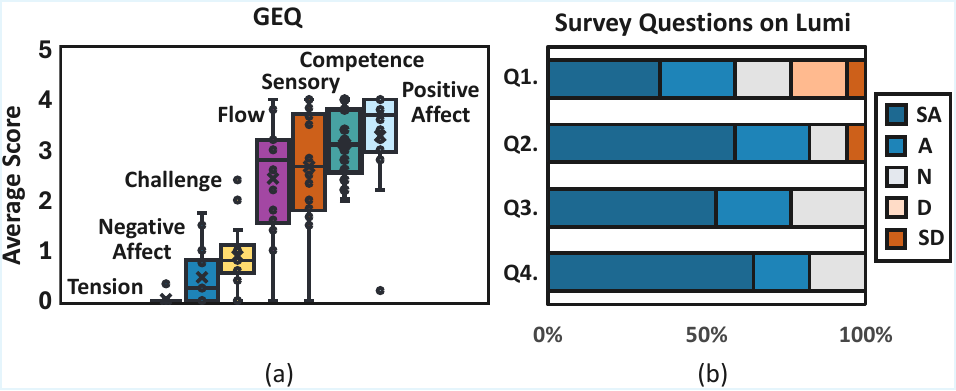}
  \vspace{-0.2cm}
  \caption{User Study II results: survey questions on \Lumi\ in 5-point Likert scale (a), and average scores for each category in GEQ (b).}
   \label{fig:study2_geq}
  \vspace{-0.45cm}
\end{figure}

\begin{figure*}[t]
  \centering
  \includegraphics[width=0.93\textwidth]{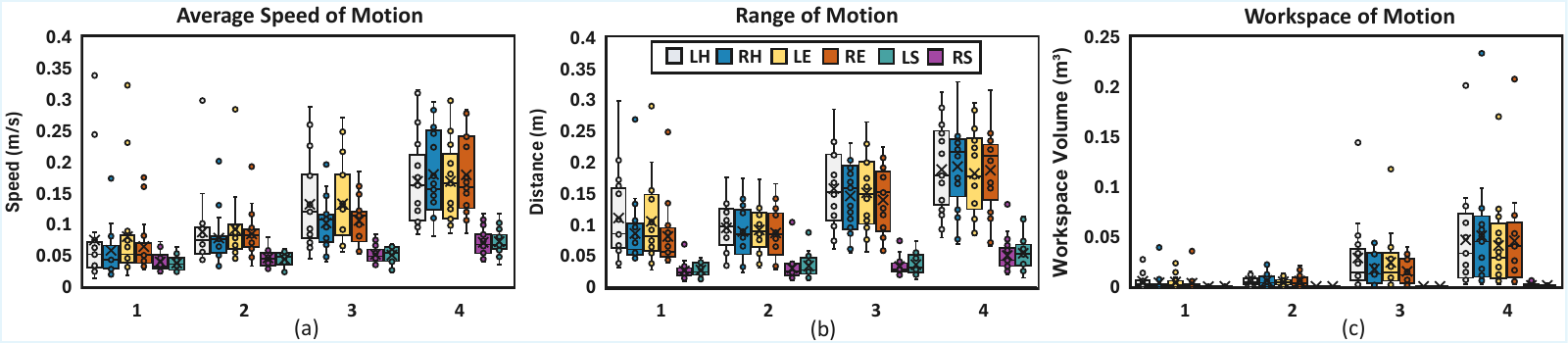}
  \vspace{-0.2cm}
  \caption{User Study II results: ICU patients' motion analysis from Quest 3 body tracking for each body joints (LH/RH: left/right hands, LE/RE: left/right elbows, LS/RS: left/right shoulders) in terms of speed (a), range of motion (b), and workspace volume (c).}
   \label{fig:study2_movement}
  \vspace{-0.4cm}
\end{figure*}

\begin{figure}[t]
  \centering
  \includegraphics[width=0.83\linewidth]{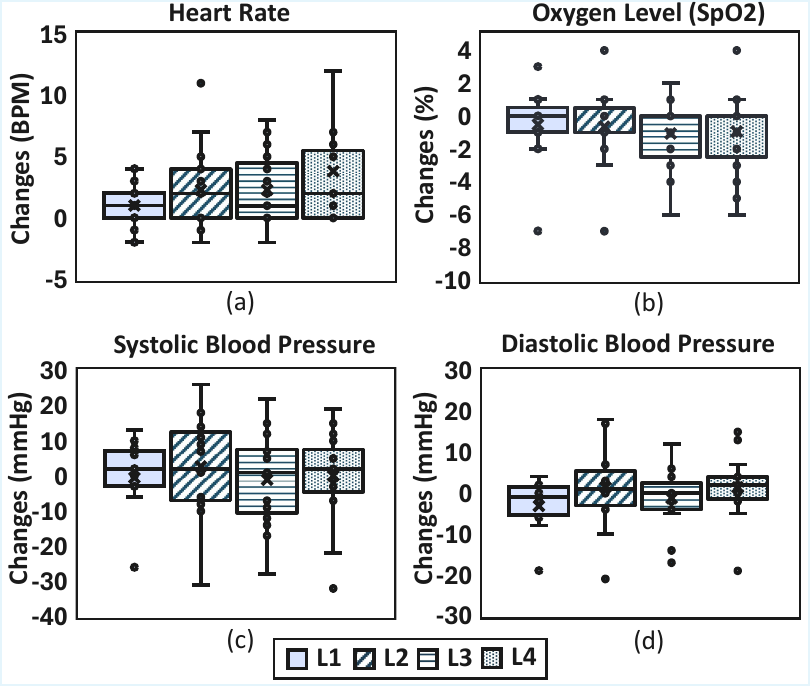}
  \vspace{-0.2cm}
  \caption{User Study II results: percentage changes of physiological measurements, compared to the baselines, including HR (a), oxygen level (b), SBP (c), and DBP (d).}
   \label{fig:study2_physiological}
  \vspace{-0.5cm}
\end{figure}

\subsection{User Study II}

\subsubsection{Game Experience}

ICU patients reported GEQ scores that were comparable to those observed in User Study I. Overall, patients demonstrated high scores in \textit{positive affect} (M = 3.18, SD = 1.03), \textit{competence} (M = 3.01, SD = 0.75), \textit{sensory and imaginative immersion} (M = 2.67, SD = 1.08), and \textit{flow} (M = 2.42, SD = 1.06), on a 0–4 scale. Conversely, low scores were reported for \textit{challenge} (M = 1.11, SD = 0.76), \textit{negative affect} (M = 0.56, SD = 0.66), and \textit{tension} (M = 0.18, SD = 0.49). These findings, which closely mirror those of healthy participants in User Study I, suggest that despite the psychological and cognitive challenges often associated with ICU-acquired weakness and prolonged immobilization, the VR exergame provided an engaging and enjoyable experience without increasing stress. Patients described the game as “calming, soft, fun to play,” (P7, P8, P17) “something different from the ICU routine,” (P3, P13) and said it made them feel “more active” (P2) and “not just lying in bed.” (P3) Several expressed interest in playing again if given more variety, asking for “new features” and “different movements to try” (P5, P9). While not all patients were able to provide detailed feedback due to communication or fatigue limitations, those who did generally endorsed the experience as positive and motivating, while suggesting more dynamic environments and interactions for future versions.

\textbf{Survey Questions on \Lumi.} Patient responses regarding \Lumi were generally favorable.~58.8\% of patients agreed that interacting with \Lumi\ resembled a social interaction, a rate comparable to User Study I (69.2\%). A higher proportion of ICU patients perceived \Lumi\ as conveying expertise (76.5\%) and providing safe guidance (82.4\%), compared to participants in User Study I (61.5\% and 69.2\%, respectively). We hypothesize that \Lumi's instructions may have been interpreted as more professional and validated within the ICU context. However, fewer ICU patients (82.4\%) endorsed \Lumi's support than in User Study I (92.3\%). While patients liked various aspects of \Lumi\ being “cheerful, happy" (P3, P7, P12), and “encouraging" (P3, P8), they also identified areas for improvement, including offering more “interaction" (P5) and “instructions" (P2). These findings suggest opportunities to improve \Lumi's design by integrating richer instructional context and expanding interactive features to strengthen perceptions of safety and support. 


\subsubsection{Physical Activity}
ICU patients in User Study II reported mean RPE (M = 2.71, SD = 4.51) and PACES (M = 101.9, SD = 16.8) scores comparable to those of healthy participants in User Study I. These results suggest that ICU patients, despite their mobility limitations, perceived the exertion as light and reported enjoyment levels comparable to the healthy participants in User Study I. We hypothesize that the light perceived exertion can be attributed to the exergame’s calibration feature, which adjusted movement boundaries to match each patient’s individual mobility capacity. As a result, even patients with critical conditions were able to complete all game levels within a personalized range of motion. This indicates that the exergame successfully avoided imposing excessive physical strain or pain while preserving high levels of engagement and ensuring task completion. 

With the exception of one excluded patient (P14), who was not sufficiently well to complete the session on the day of participation (see Section~\ref{sec:study_results}), no other participants reported nausea or motion sickness associated with VR gameplay. This absence of cybersickness is likely due to the stationary design of the exergame, with all interactions presented directly in front of the player and without requiring large-scale movements or frequent head rotations, minimizing visual-vestibular conflict. Future versions of exergames that introduce greater physical demands or broader movement requirements will require careful assessment of cybersickness in ICU patients.

\subsubsection{Physiological Changes}
HR increased gradually across game levels with progressively more challenging mobility tasks (Fig. \ref{fig:study2_physiological}a). Relative to baseline, HR increased by 1.06 BPM (SD = 1.89) after L1, 2.12 BPM (SD = 3.88) after L2, 2.12 BPM (SD = 2.55) after L3, and 4.00 BPM (SD = 6.62) after L4. ANOVA revealed a significant main effect of game level on HR, $F(4, 64) = 4.42,\ p = .003$, corroborated by the Friedman test, $\chi^2(4) = 20.05,\ p < .001,\ W = 0.29$. These increases were smaller in magnitude compared to the healthy participant group in User Study I. On the other hand, oxygen saturation showed a gradual decline across levels (–0.59\% after L1, –0.71\% after L2, –1.18\% after L3, –1.06\% after L4), but these changes were not statistically significant (ANOVA: $F(4, 64) = 0.93,\ p = .451$; Friedman: $\chi^2(4) = 3.49,\ p = .479,\ W = 0.05$).

Blood pressure exhibited upward trends with increasing task difficulty. SBP rose by 0.53 mmHg (SD = 10.68) after L1, 2.71 mmHg (SD = 13.20) after L2, -1.06 mmHg (SD = 13.06) after L3, and 0.12 mmHg (SD = 12.57) after L4, but this trend did not reach statistical significance (ANOVA: $F(4, 64) = 0.52,\ p = .724$; Friedman: $\chi^2(4) = 3.58,\ p = .466$). By contrast, DBP increased significantly (–3.18 mmHg after L1, +1.47 mmHg after L2, -1.24 mmHg after L3, +1.41 mmHg after L4), with ANOVA confirming a main effect of level, $F(4, 64) = 2.78,\ p = .034$, and Friedman test corroborating this result, $\chi^2(4) = 9.64,\ p = .047$. 

Post-hoc comparisons indicated that HR increases were gradual, with no individual level-to-level difference reaching significance after Bonferroni correction (all corrected $p > .05$), suggesting cumulative rather than stepwise changes. However, other comparisons were non-significant after correction. These results indicate that the exergame elicited measurable cardiovascular engagement in ICU patients, primarily reflected in HR, while maintaining stable oxygen saturation and without inducing unsafe physiological responses.  


\subsubsection{Movement Analysis}

Mean speed of motion increased steadily from L1 to L4 (Fig. \ref{fig:study2_movement}a), most prominently in the hands and forearms (e.g., LH: 0.075 to 0.169 m/s; RH: 0.058 to 0.180 m/s; LE: 0.080 to 0.169 m/s; RE: 0.064 to 0.179 m/s). ANOVA confirmed significant effects of level for all four joints (LH: $F(3,48)=27.89$, RH: $F=36.49$, LE: $F=27.18$, RE: $F=38.54$, all $p<.001$). Post-hoc comparisons indicated that most increases occurred from L2 onward, with smaller differences between L1 and L2. Shoulder joints also showed significant but more modest gains (LS: 0.040–0.071 m/s, $F(3,48)=31.15$; RS: 0.038–0.072 m/s, $F(3,48)=38.99$, both $p<.001$), with significance concentrated in late-level contrasts (L1-L3, L1–L4, L2–L4, L3–L4). 

ROM increased systematically, particularly in the hands and forearms, while shoulders again showed significance mainly at later levels (Fig. \ref{fig:study2_movement}b). Distal joints expanded from 0.09–0.11m at L1–L2 to nearly 0.19m at L4 (LH = 0.188m, RH = 0.193m, LE = 0.182m, RE = 0.187m). ANOVA confirmed significant effects (LH: $F(3,48)=12.13$, RH: $F=18.10$, LE: $F=11.87$, RE: $F=19.05$, all $p<.001$), and post-hoc tests indicated significant changes across most level pairs except the earliest (L1–L2, L1–L3). The ROM of shoulder joints remained relatively stable through L3 (0.026–0.036m) but increased at L4 (LS = 0.051m, RS = 0.054m), with ANOVA confirming significant effects (LS: $F(3,48)=10.17$, RS: $F=14.60$, both $p<.001$) and post-hoc significance limited to late-level contrasts.

Workspace volume also expanded with increasing level, especially for distal joints (hands and elbows), though absolute values were smaller than those of healthy participants (Fig. \ref{fig:study2_movement}c). By L4, distal joints reached volumes of 0.041–0.051 $m^3$, while proximal (shoulders) remained below 0.002 $m^3$. ANOVA confirmed significant effects for hands and forearms (LH: $F(3,48)=10.82$, RH: $F=10.45$, LE: $F=11.43$, RE: $F=10.68$, all $p<.001$), with post-hoc comparisons showing increases primarily between early and late levels (L1–L4, L2–L4). Shoulder effects were weaker (LS: $F(3,48)=8.07$, RS: $F=8.40$, both $p<.001$), with only L3–L4 or late-level contrasts reaching significance.

Overall, ICU patients demonstrated the same systematic increases in speed, ROM, and workspace as healthy participants in User Study I, validating the scalability of the exergame. However, absolute performance remained substantially lower, reflecting their physical limitations, and the disparity between distal and proximal joints was more pronounced. This pattern suggests that while the exergame effectively engages ICU patients across all joints, rehabilitation gains are likely to be most pronounced in distal movements, highlighting the importance of tailoring progression and support for proximal joint recovery. These metrics could enable physical therapists to track progress across sessions, identify limitations between distal and proximal mobility, and adapt therapy plans accordingly. In future work, we plan to collaborate with clinicians to define how such quantitative measures can complement existing assessment tools and support long-term recovery monitoring.

\subsubsection{Survey-based Feedback from the Care Team}

Seven expert nurses completed the 
survey after conducting 18 sessions with ICU patients. Overall, the care team reported that the system was quick to set up, easy to operate, and straightforward to follow. They noted that the VR system functioned reliably with no major technical issues, aside from occasional delays in body tracking when switching between the controller and hand-tracking modes. However, they suggested improvements to hand-tracking accuracy and expressed concern that VR experience limited patients' visibility of surrounding medical equipment. The team also suggested that \Lumi’s instructions could be made clearer and that incorporating a more dynamic environment and varied music would improve engagement, particularly for patients using the system on consecutive days. Importantly, \textit{the nurses strongly agreed that the sessions met the intended therapeutic goals}, were effective and productive, and endorsed the use of VR technology in the ICU setting without reservations.

\section{Discussion} \label{sec:discussion}

\subsection{System Usability in ICU Setting} \label{sec:discussion_system_usability}

In critical care, the adoption of new rehabilitation tools depends not only on patient acceptance and outcomes but also on clinical usability and workflow integration. Our findings suggest that \textit{the VR exergame was sufficiently intuitive for clinicians with minimal prior VR experience}, highlighting its potential for use in the time-sensitive and resource-constrained ICU environment. Open-ended feedback aligned with this observation, with clinicians describing the system as “easy to navigate" and that “instructions were clear", while others praised our exergame as an “excellent addition to physical rehabilitation" with “huge potential for ICU patients." At the same time, they identified practical barriers, including the need to streamline casting, improve the FoV on casted displays, and reduce reliance on handheld controllers. These challenges have similarly been noted in other VR rehabilitation studies, where technical setup and interface complexity are known to hinder adoption in clinical settings~\cite{perez2022virtual, woodbridge2023clinician}. Addressing such issues will be important for future refinements to ensure seamless integration into ICU workflows.


\subsection{Validation of Exergame Design \& Experience} \label{sec:discussion_design_validation}

\textbf{Engagement and Game Experience.} GEQ, PACES, and RPE results indicate that the exergame consistently generated enjoyment and engagement, even among ICU patients who may experience cognitive impairment from ICU-acquired weakness. 
This is notable in critical care, where conventional rehabilitation often faces poor adherence due to discomfort or low motivation~\cite{woodbridge2023clinician}. Prior work has shown that VR can improve patient acceptance of early mobilization in the ICU by enhancing engagement and reducing perceived exertion~\cite{de2025feasibility}. Our study extends this evidence by not only embedding progressive mobility challenges, but also by systematically monitoring patient experiences and movement data with Quest 3, and by demonstrating that the exergame improved mobility while maintaining safe physiological responses.


The character, \emph{\Lumi}, played a key role in shaping the experiential quality of the intervention. ICU patients reported that \Lumi\ enhanced feelings of safety and support, and conveyed a stronger sense of expertise compared to the responses from healthy participants, likely reflecting ICU patients’ reliance on healthcare professionals for guidance. Similar effects have been reported in prior work, where virtual agents displaying empathy and authority improved trust and adherence \cite{tironi2019empathic, graf2023emotional}. Our findings extend these insights to the ICU, highlighting the value of designing agents that communicate both empathy and expertise to foster engagement and therapeutic credibility. 

\textbf{Mobility and Physiological Safety.} Beyond subjective experiences, the exergame elicited progressive increases in upper-body mobility, including greater joint movement speed, expanded range of motion, and larger workspace volume across game levels. These kinematic improvements demonstrate that \textit{the game successfully elicited purposeful motor activity rather than passive engagement}, thereby validating its design as a movement-oriented rehabilitation tool. For ICU patients at risk of muscle atrophy, joint stiffness, and loss of functional capacity, these improvements represent clinically meaningful engagement with early mobilization protocols.  

Physiological measurements further underscored the appropriateness of the exergame for use in critical care. Whereas healthy participants showed expected cardiovascular responses, ICU patients exhibited smaller fluctuations in heart rate, blood pressure, and oxygen saturation. This attenuation likely reflects masked physiological responses of post-surgical ICU care, where medications \cite{van2019metabolic} and hemodynamic stabilizers can dampen autonomic variability \cite{pinsky2007hemodynamic}. These results indicate that the exergame activated motor pathways without imposing excessive cardiovascular stress, validating its safety and feasibility for ICU patients.  

\subsection{Limitations and Future Work} \label{sec:discussion_implementation}
This work focuses on the feasibility and clinical acceptance of VR exergames in the ICU. Although Quest 3 body tracking demonstrated sufficient accuracy for the evaluated tasks, its inferred joint estimation reflects an inherent trade-off between tracking precision and deployability in critical care settings. As future work, we plan to explore the integration of lightweight wearable trackers, such as the VIVE Ultimate Tracker, to support more complex or fine-grained mobility challenges 
In addition, while the current exergame employs predefined mobility levels personalized through patient-specific boundary calibration, future iterations will investigate adaptive progression based on patients’ real-time performance and physiological signals to enable more responsive and individualized rehabilitation. Finally, as the primary goal of this work was to establish the feasibility, safety, and clinical acceptance of VR-based early mobilization in a highly constrained critical care environment, we did not include a control group or longitudinal follow-up. We are working on expanding the game levels to a more complex and broader range of motion to assess longer-term functional outcomes, recovery trajectories, and integration with standard physical therapy workflows, particularly for ICU patients with extended lengths of stay.

\section{Conclusion}
\label{sec:conclusion}

This work introduced a patient-centered VR exergame designed to support early upper-body mobilization of ICU patients, addressing both therapeutic and clinical workflow needs. Through two studies, first with healthy participants and healthcare professionals or students in a simulated setting, and later with cardiothoracic ICU patients and a dedicated care team, we demonstrated the system’s feasibility, usability, and clinical acceptance. Our results show that the exergame delivered engaging game experiences, elicited measurable improvements in mobility, and maintained physiological safety in a critical care environment. These findings validate the potential of VR exergames as an adjunct to conventional rehabilitation practices in ICUs.


\acknowledgments{
We thank all nurses and patients who participated in our study. This work was supported in part by NSF grants CSR-2312760, CNS-2112562 and IIS-2231975, NSF CAREER Award IIS-2046072, NSF NAIAD Award 2332744, and by Duke University School of Nursing Center for Nursing Research.}

\bibliographystyle{abbrv-doi-hyperref}

\bibliography{template}

\begin{thebibliography}{10}

\bibitem{abdlkarim2024methodological}
D.~Abdlkarim, M.~Di~Luca, P.~Aves, M.~Maaroufi, S.-H. Yeo, R.~C. Miall, P.~Holland, and J.~M. Galea.
\newblock A methodological framework to assess the accuracy of virtual reality hand-tracking systems: {A} case study with the {Meta Quest 2}.
\newblock {\em Behavior Research Methods}, 56(2):1052--1063, 2024.

\bibitem{afsar2018virtual}
S.~I. Afsar, I.~Mirzayev, O.~U. Yemisci, and S.~N.~C. Saracgil.
\newblock Virtual reality in upper extremity rehabilitation of stroke patients: {A} randomized controlled trial.
\newblock {\em Journal of Stroke and Cerebrovascular Diseases}, 27(12):3473--3478, 2018.

\bibitem{allen2007validity}
D.~D. Allen.
\newblock Validity and reliability of the movement ability measure: {A} self-report instrument proposed for assessing movement across diagnoses and ability levels.
\newblock {\em Physical Therapy}, 87(7):899--916, 2007.

\bibitem{bartl2022effects}
A.~Bartl, C.~Merz, D.~Roth, and M.~E. Latoschik.
\newblock The effects of avatar and environment design on embodiment, presence, activation, and task load in a virtual reality exercise application.
\newblock In {\em Proceedings of the IEEE International Symposium on Mixed and Augmented Reality (ISMAR)}, pp. 260--269, 2022.

\bibitem{brazil2024effect}
C.~K. Brazil and M.~J. Rys.
\newblock {The effect of VR on fine motor performance by older adults: A comparison between real and virtual tasks}.
\newblock {\em Virtual Reality}, 28(2):113, 2024.

\bibitem{brooke1996sus}
J.~Brooke et~al.
\newblock {SUS-A} quick and dirty usability scale.
\newblock {\em Usability Evaluation in Industry}, 189(194):4--7, 1996.

\bibitem{canning2020virtual}
C.~G. Canning, N.~E. Allen, E.~Nackaerts, S.~S. Paul, A.~Nieuwboer, and M.~Gilat.
\newblock Virtual reality in research and rehabilitation of gait and balance in {P}arkinson disease.
\newblock {\em Nature Reviews Neurology}, 16(8):409--425, 2020.

\bibitem{capecci2019kimore}
M.~Capecci, M.~G. Ceravolo, F.~Ferracuti, S.~Iarlori, A.~Monteriu, L.~Romeo, and F.~Verdini.
\newblock The {KIMORE} dataset: {K}inematic assessment of movement and clinical scores for remote monitoring of physical rehabilitation.
\newblock {\em IEEE Transactions on Neural Systems and Rehabilitation Engineering}, 27(7):1436--1448, 2019.

\bibitem{caserman2019real}
P.~Caserman, A.~Garcia-Agundez, R.~Konrad, S.~G{\"o}bel, and R.~Steinmetz.
\newblock Real-time body tracking in virtual reality using a {Vive} tracker.
\newblock {\em Virtual Reality}, 23(2):155--168, 2019.

\bibitem{casile2023quantitative}
A.~Casile, G.~Fregna, V.~Boarini, C.~Paoluzzi, F.~Manfredini, N.~Lamberti, A.~Baroni, and S.~Straudi.
\newblock Quantitative comparison of hand kinematics measured with a markerless commercial head-mounted display and a marker-based motion capture system in stroke survivors.
\newblock {\em Sensors}, 23(18):7906, 2023.

\bibitem{chen2020immersive}
W.~Chen, M.~Bang, D.~Krivonos, H.~Schimek, and A.~Naval.
\newblock {An immersive virtual reality exergame for people with Parkinson’s disease}.
\newblock In {\em International Conference on Computers Helping People with Special Needs}, pp. 138--145. Springer, 2020.

\bibitem{cmentowski2023never}
S.~Cmentowski, S.~Karaosmanoglu, L.~E. Nacke, F.~Steinicke, and J.~H. Kr{\"u}ger.
\newblock Never skip leg day again: {T}raining the lower body with vertical jumps in a virtual reality exergame.
\newblock In {\em Proceedings of the CHI Conference on Human Factors in Computing Systems}, pp. 1--18, 2023.

\bibitem{craig2003international}
C.~L. Craig, A.~L. Marshall, M.~Sj{\"o}str{\"o}m, A.~E. Bauman, M.~L. Booth, B.~E. Ainsworth, M.~Pratt, U.~Ekelund, A.~Yngve, J.~F. Sallis, et~al.
\newblock International physical activity questionnaire: 12-country reliability and validity.
\newblock {\em Medicine \& Science in Sports \& Exercise}, 35(8):1381--1395, 2003.

\bibitem{de2025feasibility}
M.~de~Vries, L.~F. Beumeler, J.~van~der Meulen, C.~Bethlehem, R.~den Otter, and E.~C. Boerma.
\newblock The feasibility of virtual reality therapy for upper extremity mobilization during and after intensive care unit admission.
\newblock {\em PeerJ}, 13:e18461, 2025.

\bibitem{duval2022designing}
J.~Duval, R.~Thakkar, D.~Du, K.~Chin, S.~Luo, A.~Elor, M.~S. El-Nasr, and M.~John.
\newblock {Designing spellcasters from clinician perspectives: A customizable gesture-based immersive virtual reality game for stroke rehabilitation}.
\newblock {\em ACM Transactions on Accessible Computing (TACCESS)}, 15(3):1--25, 2022.

\bibitem{elor2018project}
A.~Elor, M.~Teodorescu, and S.~Kurniawan.
\newblock {Project Star Catcher: A novel immersive virtual reality experience for upper limb rehabilitation}.
\newblock {\em ACM Transactions on Accessible Computing (TACCESS)}, 11(4):1--25, 2018.

\bibitem{eom2025legato}
S.~Eom, W.~Xu, L.~Zou, A.~Frith, E.~Escobar, G.~Streisfeld, A.~Mall, B.~Granger, and M.~Gorlatova.
\newblock Legato: {V}irtual reality for physical rehabilitation of patients in the intensive care unit.
\newblock In {\em Proceedings of the IEEE Conference on Virtual Reality and 3D User Interfaces Abstracts and Workshops (VRW)}, pp. 1027--1032, 2025.

\bibitem{fu2022systematic}
Y.~Fu, Y.~Hu, and V.~Sundstedt.
\newblock A systematic literature review of virtual, augmented, and mixed reality game applications in healthcare.
\newblock {\em ACM Transactions on Computing for Healthcare (HEALTH)}, 3(2):1--27, 2022.

\bibitem{godden2025robotic}
E.~Godden, W.~Steedman, and M.~K. Pan.
\newblock Robotic characterization of markerless hand-tracking on {Meta Quest Pro and Quest 3} virtual reality headsets.
\newblock {\em IEEE Transactions on Visualization and Computer Graphics}, 2025.

\bibitem{graf2023emotional}
L.~Graf, S.~Abramowski, F.~Born, and M.~Masuch.
\newblock Emotional virtual characters for improving motivation and performance in {VR} exergames.
\newblock {\em Proceedings of the ACM on Human-Computer Interaction}, 7(CHI PLAY):1115--1135, 2023.

\bibitem{grupp2017evo}
M.~Grupp.
\newblock {evo: Python package for the evaluation of odometry and SLAM}, 2017.

\bibitem{hamzeheinejad2021impact}
N.~Hamzeheinejad, D.~Roth, S.~Monty, J.~Breuer, A.~Rodenberg, and M.~E. Latoschik.
\newblock The impact of implicit and explicit feedback on performance and experience during {VR}-supported motor rehabilitation, 2021.

\bibitem{he2025effects}
Y.~He, Q.~Yang, X.~Dai, T.~Chen, H.~Wu, K.~Li, S.~Zhu, Y.~Liu, and H.~Lei.
\newblock Effects of virtual reality technology on early mobility in critically ill adult patients: {A} systematic review and meta-analysis.
\newblock {\em Frontiers in Neurology}, 15:1469079, 2025.

\bibitem{howard2017meta}
M.~C. Howard.
\newblock A meta-analysis and systematic literature review of virtual reality rehabilitation programs.
\newblock {\em Computers in Human Behavior}, 70:317--327, 2017.

\bibitem{hu2024apple}
T.~Hu, F.~Yang, T.~Scargill, and M.~Gorlatova.
\newblock {Apple vs Meta: A} comparative study on spatial tracking in {SOTA XR} headsets.
\newblock In {\em Proceedings of the International Conference on Mobile Computing and Networking}, pp. 2120--2127, 2024.

\bibitem{ijsselsteijn2013game}
W.~A. IJsselsteijn, Y.~A. De~Kort, and K.~Poels.
\newblock The game experience questionnaire.
\newblock 2013.

\bibitem{jacob2021multidisciplinary}
P.~Jacob, P.~Gupta, S.~Shiju, A.~S. Omar, S.~Ansari, G.~Mathew, M.~Varghese, J.~Pulimoottil, S.~Varkey, M.~Mahinay, et~al.
\newblock {Multidisciplinary, early mobility approach to enhance functional independence in patients admitted to a cardiothoracic intensive care unit: A quality improvement programme}.
\newblock {\em BMJ Open Quality}, 10(3):e001256, 2021.

\bibitem{jang2019pulmonary}
M.~H. Jang, M.-J. Shin, and Y.~B. Shin.
\newblock Pulmonary and physical rehabilitation in critically ill patients.
\newblock {\em Acute and Critical Care}, 34(1):1, 2019.

\bibitem{kaminska2022usability}
D.~Kami{\'n}ska, G.~Zwoli{\'n}ski, and A.~Laska-Le{\'s}niewicz.
\newblock Usability testing of virtual reality applications—the pilot study.
\newblock {\em Sensors}, 22(4):1342, 2022.

\bibitem{karageorghis2011revisiting}
C.~I. Karageorghis, L.~Jones, D.-L. Priest, R.~I. Akers, A.~Clarke, J.~M. Perry, B.~T. Reddick, D.~T. Bishop, and H.~B. Lim.
\newblock Revisiting the relationship between exercise heart rate and music tempo preference.
\newblock {\em Research Quarterly for Exercise and Sport}, 82(2):274--284, 2011.

\bibitem{karageorghis2012music}
C.~I. Karageorghis and D.-L. Priest.
\newblock {Music in the exercise domain: A review and synthesis (Part I)}.
\newblock {\em International Review of Sport and Exercise Psychology}, 5(1):44--66, 2012.

\bibitem{kendzierski1991physical}
D.~Kendzierski and K.~J. DeCarlo.
\newblock {Physical activity enjoyment scale: Two validation studies}.
\newblock {\em Journal of Sport and Exercise Psychology}, 13(1):50--64, 1991.

\bibitem{kern2019immersive}
F.~Kern, C.~Winter, D.~Gall, I.~K{\"a}thner, P.~Pauli, and M.~E. Latoschik.
\newblock Immersive virtual reality and gamification within procedurally generated environments to increase motivation during gait rehabilitation.
\newblock In {\em Proceedings of the IEEE Conference on Virtual Reality and 3D User Interfaces (VR)}, pp. 500--509, 2019.

\bibitem{laver2018virtual}
K.~E. Laver, B.~Lange, S.~George, J.~E. Deutsch, G.~Saposnik, and M.~Crotty.
\newblock Virtual reality for stroke rehabilitation.
\newblock {\em Stroke}, 49(4):e160--e161, 2018.

\bibitem{li2024role}
M.~Li, M.~Race, F.~Huang, and M.~X. Escalon.
\newblock The role of virtual reality to promote mobilization in the critical care setting: {A} narrative review.
\newblock {\em American Journal of Physical Medicine \& Rehabilitation}, pp. 10--1097, 2024.

\bibitem{liao2025focus}
K.-L. Liao, M.~Huang, J.~Shi, M.~Chen, and R.~Yang.
\newblock Focus-driven augmented feedback: {E}nhancing focus and maintaining engagement in upper limb virtual reality rehabilitation.
\newblock {\em IEEE Transactions on Visualization and Computer Graphics}, 2025.

\bibitem{linke2020early}
C.~A. Linke, L.~B. Chapman, L.~J. Berger, T.~L. Kelly, C.~A. Korpela, and M.~G. Petty.
\newblock {Early mobilization in the ICU: A collaborative, integrated approach}.
\newblock {\em Critical Care Explorations}, 2(4):e0090, 2020.

\bibitem{lourido2024using}
C.~Lourido, Z.~Waghoo, H.~K. Wazir, N.~Bhagat, and V.~Kapila.
\newblock Using capability maps tailored to arm range of motion in {VR} exergames for rehabilitation.
\newblock In {\em Proceedings of the IEEE International Conference of the Engineering in Medicine and Biology Society (EMBC)}, pp. 1--4, 2024.

\bibitem{mall2024virtual}
A.~Mall, J.~Stokes, G.~Streisfeld, M.~Zychowicz, and B.~B. Granger.
\newblock Virtual reality strategies for promoting mobility in the intensive care unit: {A} case report.
\newblock {\em AACN Advanced Critical Care}, 35(3):238--243, 2024.

\bibitem{marotta2022integrating}
N.~Marotta, D.~Calafiore, C.~Curci, L.~Lippi, V.~Ammendolia, F.~Ferraro, M.~Invernizzi, et~al.
\newblock {Integrating virtual reality and exergaming in cognitive rehabilitation of patients with Parkinson disease: A systematic review of randomized controlled trials}.
\newblock {\em European Journal of Physical and Rehabilitation Medicine}, 58(6):818, 2022.

\bibitem{mc2023patient}
A.~Mc~Kittrick, M.~R. Desselle, A.~Padilha Lanari~Bo, B.~Zhang, S.~Laracy, and G.~Tornatore.
\newblock Patient experience in adjunct controller-free hand tracking virtual reality tasks for upper-limb occupational therapy rehabilitation.
\newblock {\em Journal of Patient Experience}, 10:23743735231211983, 2023.

\bibitem{metabodytracking}
{Meta Horizon}.
\newblock Movement body tracking {OpenXR} extension.
\newblock \url{https://developers.meta.com/horizon/documentation/unity/move-body-tracking/}, 2025.

\bibitem{metapassthrough}
{Meta Horizon}.
\newblock Passthrough camera {API} overview.
\newblock \url{https://developers.meta.com/horizon/documentation/spatial-sdk/spatial-sdk-pca-overview/}, 2025.

\bibitem{metaunitysentis}
{Meta Horizon}.
\newblock Unity {Sentis} for on-device {ML/CV} models.
\newblock \url{https://developers.meta.com/horizon/documentation/unity/unity-pca-sentis/}, 2025.

\bibitem{miller2022temporal}
R.~Miller, N.~K. Banerjee, and S.~Banerjee.
\newblock Temporal effects in motion behavior for virtual reality ({VR}) biometrics.
\newblock In {\em Proceedings of the IEEE Conference on Virtual Reality and 3D User Interfaces (VR)}, pp. 563--572, 2022.

\bibitem{minimax}
Minimax.
\newblock Minimax audio.
\newblock https://www.minimax.io/audio, 2025.

\bibitem{palaniappan2018developing}
S.~M. Palaniappan and B.~S. Duerstock.
\newblock Developing rehabilitation practices using virtual reality exergaming.
\newblock In {\em Proceedings of the IEEE International Symposium on Signal Processing and Information Technology (ISSPIT)}, pp. 090--094, 2018.

\bibitem{perez2022virtual}
V.~Z. P{\'e}rez, J.~C. Yepes, J.~F. Vargas, J.~C. Franco, N.~I. Escobar, L.~Betancur, J.~S{\'a}nchez, and M.~J. Betancur.
\newblock Virtual reality game for physical and emotional rehabilitation of landmine victims.
\newblock {\em Sensors}, 22(15):5602, 2022.

\bibitem{pinsky2007hemodynamic}
M.~R. Pinsky.
\newblock Hemodynamic evaluation and monitoring in the {ICU}.
\newblock {\em Chest}, 132(6):2020--2029, 2007.

\bibitem{reinhardt2019entropy}
D.~Reinhardt, S.~Haesler, J.~Hurtienne, and C.~Wienrich.
\newblock Entropy of controller movements reflects mental workload in virtual reality.
\newblock In {\em Proceedings of the IEEE Conference on Virtual Reality and 3D User Interfaces (VR)}, pp. 802--808, 2019.

\bibitem{rojo2023pedaleovr}
A.~Rojo, A.~Castrillo, C.~L{\'o}pez, L.~Perea, F.~Alnajjar, J.~C. Moreno, and R.~Raya.
\newblock {PedaleoVR: Usability study of a virtual reality application for cycling exercise in patients with lower limb disorders and elderly people}.
\newblock {\em Plos One}, 18(2):e0280743, 2023.

\bibitem{rose2018immersion}
T.~Rose, C.~S. Nam, and K.~B. Chen.
\newblock Immersion of virtual reality for rehabilitation-review.
\newblock {\em Applied Ergonomics}, 69:153--161, 2018.

\bibitem{saha2022mapping}
S.~Saha, H.~Noble, A.~Xyrichis, D.~Hadfield, T.~Best, P.~Hopkins, and L.~Rose.
\newblock {Mapping the impact of ICU design on patients, families and the ICU team: A scoping review}.
\newblock {\em Journal of Critical Care}, 67:3--13, 2022.

\bibitem{shah2023social}
S.~H.~H. Shah, A.~S.~T. Karlsen, M.~Solberg, and I.~A. Hameed.
\newblock {A social VR-based collaborative exergame for rehabilitation: Codesign, development and user study}.
\newblock {\em Virtual Reality}, 27(4):3403--3420, 2023.

\bibitem{singam2024mobilizing}
A.~Singam.
\newblock Mobilizing progress: {A} comprehensive review of the efficacy of early mobilization therapy in the intensive care unit.
\newblock {\em Cureus}, 16(4), 2024.

\bibitem{skuric2022approximating}
A.~Skuric, V.~Padois, and D.~Daney.
\newblock Approximating robot reachable space using convex polytopes.
\newblock In {\em International Workshop on Human-Friendly Robotics}, pp. 45--60. Springer, 2022.

\bibitem{soderberg2022fear}
A.~S{\"o}derberg, V.~Karlsson, B.~M. Ahlberg, A.~Johansson, and A.~Thelandersson.
\newblock From fear to fight: {P}atients experiences of early mobilization in intensive care. a qualitative interview study.
\newblock {\em Physiotherapy Theory and Practice}, 38(6):750--758, 2022.

\bibitem{tironi2019empathic}
A.~Tironi, R.~Mainetti, M.~Pezzera, and N.~A. Borghese.
\newblock An empathic virtual caregiver for assistance in exer-game-based rehabilitation therapies.
\newblock In {\em Proceedings of the IEEE International Conference on Serious Games and Applications for Health (SeGAH)}, pp. 1--6, 2019.

\bibitem{van2019metabolic}
C.~H. van Herpen, D.~A. van Blokland, and A.~R. van Zanten.
\newblock Metabolic effects of beta-blockers in critically ill patients: {A} retrospective cohort study.
\newblock {\em Heart \& Lung}, 48(4):278--286, 2019.

\bibitem{vanhorebeek2020icu}
I.~Vanhorebeek, N.~Latronico, and G.~Van~den Berghe.
\newblock {ICU}-acquired weakness.
\newblock {\em Intensive Care Medicine}, 46(4):637--653, 2020.

\bibitem{vlake2022intensive}
J.~H. Vlake, J.~van Bommel, E.-J. Wils, J.~Bienvenu, M.~E. Hellemons, T.~I. Korevaar, A.~F. Schut, J.~A. Labout, L.~L. Schreuder, M.~P. van Bavel, et~al.
\newblock {Intensive care unit--specific virtual reality for critically ill patients with COVID-19: Multicenter randomized controlled trial}.
\newblock {\em Journal of Medical Internet Research}, 24(1):e32368, 2022.

\bibitem{wang2022survey}
L.~Wang, M.~Huang, R.~Yang, H.-N. Liang, J.~Han, and Y.~Sun.
\newblock Survey of movement reproduction in immersive virtual rehabilitation.
\newblock {\em IEEE Transactions on Visualization and Computer Graphics}, 29(4):2184--2202, 2022.

\bibitem{wang2022supporting}
Q.~Wang, B.~Kang, and P.~O. Kristensson.
\newblock Supporting playful rehabilitation in the home using virtual reality headsets and force feedback gloves.
\newblock In {\em Proceedings of the IEEE Conference on Virtual Reality and 3D User Interfaces (VR)}, pp. 504--513, 2022.

\bibitem{wang2020early}
T.-H. Wang.
\newblock Early mobilization on patients with mechanical.
\newblock {\em Physical Therapy Effectiveness}, p. 193, 2020.

\bibitem{wang2024quantitative}
X.~Wang, J.~Zhang, S.~Q. Xie, C.~Shi, J.~Li, and Z.-Q. Zhang.
\newblock Quantitative upper limb impairment assessment for stroke rehabilitation: {A} review.
\newblock {\em IEEE Sensors Journal}, 24(6):7432--7447, 2024.

\bibitem{williams2017borg}
N.~Williams.
\newblock {The Borg rating of perceived exertion (RPE) scale}.
\newblock {\em Occupational Medicine}, 67(5):404--405, 2017.

\bibitem{woodbridge2023clinician}
H.~R. Woodbridge, C.~Norton, M.~Jones, S.~J. Brett, C.~M. Alexander, and A.~C. Gordon.
\newblock Clinician and patient perspectives on the barriers and facilitators to physical rehabilitation in intensive care: {A} qualitative interview study.
\newblock {\em BMJ Open}, 13(11):e073061, 2023.

\bibitem{zhang2015objective}
Z.~Zhang, Q.~Fang, and X.~Gu.
\newblock Objective assessment of upper-limb mobility for poststroke rehabilitation.
\newblock {\em IEEE Transactions on Biomedical Engineering}, 63(4):859--868, 2015.

\bibitem{ziyaeifard2018effects}
M.~Ziyaeifard, F.~G.~B. Khoo, S.~Lotfian, R.~Azarfarin, R.~Aminnejad, R.~Alikhani, and M.~Y. Moghadam.
\newblock Effects of early mobilization protocol on cognitive outcome after cardiac surgery.
\newblock {\em Annals of Anesthesiology and Critical Care}, 3(1):1--8, 2018.

\end{thebibliography}

\appendix 







\end{document}